\documentclass[12pt, letter]{emulateapj}
\usepackage{epsf}
\usepackage{graphicx,amssymb,amsmath}

\newcommand{\photoz}{photo-$z$}
\newcommand{\photozs}{photo-$z$s}

\newcommand{\msun}{M_{\odot}}
\newcommand{\sersic}{S\'{e}rsic}
\newcommand{\synmag}{{\sc synmag}}

\shorttitle{{\synmag}s: Catalog-Level Matched Photometry}
\shortauthors{Bundy et al.}

\submitted{Published in AJ, 2012, 144, 188}

\begin{document}

\title{SYNMAG Photometry: A Fast Tool for Catalog-Level Matched Colors of Extended Sources}

\author{Kevin Bundy\altaffilmark{1}, David W.~Hogg\altaffilmark{2}, Tim D.~Higgs\altaffilmark{3}, Robert
  C.~Nichol\altaffilmark{3}, Naoki Yasuda\altaffilmark{1}, Karen L.~Masters\altaffilmark{3}, Dustin Lang\altaffilmark{4}, David A.~Wake\altaffilmark{5}}


\altaffiltext{1}{Kavli Institute for the Physics and Mathematics of the Universe, Todai Institutes for Advanced
  Study, the University of Tokyo, Kashiwa, Japan 277-8583 (Kavli IPMU, WPI)}
\altaffiltext{2}{Center for Cosmology and Particle Physics, Department of Physics, New York University, 4 Washington Place, New York, NY 10003}
\altaffiltext{3}{Institute for Cosmology and Gravitation, Dennis Sciama Building, University of Portsmouth, Burnaby Road, Portsmouth PO1 3FX}
\altaffiltext{4}{Department of Astrophysical Sciences, Princeton University, Princeton, NJ 08544, USA}
\altaffiltext{5}{Department of Astronomy, Yale University, New Haven, CT 06520-8101, USA}

\begin{abstract}

  Obtaining reliable, matched photometry for galaxies imaged by different observatories represents a key challenge in the era of
  wide-field surveys spanning more than several hundred square degrees.  Methods such as flux fitting, profile fitting, and PSF
  homogenization followed by matched-aperture photometry are all computationally expensive.  We present an alternative solution
  called ``synthetic aperture photometry'' that exploits galaxy profile fits in one band to efficiently model the observed,
  PSF-convolved light profile in other bands and predict the flux in arbitrarily sized apertures.  Because aperture magnitudes are
  the most widely tabulated flux measurements in survey catalogs, producing synthetic aperture magnitudes ({\synmag}s) enables
  very fast matched photometry at the catalog level, without reprocessing imaging data.  We make our code public and apply it to
  obtain matched photometry between SDSS $ugriz$ and UKIDSS $YJHK$ imaging, recovering red-sequence colors and photometric
  redshifts with a scatter and accuracy as good as if not better than FWHM-homogenized photometry from the GAMA Survey.  Finally,
  we list some specific measurements that upcoming surveys could make available to facilitate and ease the use of {\synmag}s.

\end{abstract}


\section{Introduction}\label{intro}

Astronomy is entering an era of truly panoramic deep imaging surveys.  Building on the legacy from the Sloan Digital Sky Survey
\citep[SDSS,][]{york00, abazajian09}, the coming years will see a rapidly expanding patchwork of overlapping imaging campaigns, with surveys like
DES\footnote{Dark Energy Survey}, HSC\footnote{Hyper-Suprime Cam Survey}, KIDS\footnote{Kilo-Degree Survey},
VIKINGS\footnote{VISTA Kilo-degree Infrared Galaxy Survey}, and eventually Euclid and LSST\footnote{Large Synoptic Sky Telescope}
providing hundreds of square degrees of new data on rapid timescales.  Fast and reliable methods for combining information from
these disparate surveys are clearly needed.

A key challenge is measuring reliable matched photometry that samples, as closely as possible, the flux emitted from the same
regions of a galaxy as observed across the many wavebands sampled by these data sets.  For galaxy studies, the goal is to obtain
accurate colors and spectral energy distributions (SEDs) over a large wavelength range in order to derive photometric redshifts
(\photozs) and physical properties like stellar mass ($M_*$) from template fitting.  But as extended sources, galaxies present
particular challenges for matched photometry because their range in size, shape, and surface brightness is not easily accounted
for by a single model or choice of aperture.

Three basic methods have developed for obtaining matched photometry for extended sources from a set of multiband images with
different seeing, scales, and other properties.  The first and perhaps most common is PSF-homogenization followed by
matched-aperture photometry.  Here, all of the overlapping images in a set of observations are convolved to the PSF of the band
with the worst seeing.  The convolved images are then registered (often with interpolation to the same pixel grid) and the flux
through a common aperture is measured on the pixel data in each band at the location of every source.  This method has been
studied in detail by \citet{hildebrandt12} who stress the value of spatially-dependent convolution kernels that are designed to
make the PSF uniform across every image in all bands.  The method has been successfully applied to a variety of recent data sets
including COSMOS \citep{capak07a} and GAMA \citep{hill11}.  The disadvantages include the loss of information resulting from the
PSF degradation, the introduction of correlated noise resulting from regridding and interpolation, and the fact that aperture flux
measurements (especially for faint sources) are not ideal flux estimators.

The second method for obtaining matched photometry is ``flux fitting.''  This approach is often adopted when the PSF sizes of the
imaging data are strongly mismatched, for example when comparing photometry from {\it HST} to ground-based imaging or much lower
resolution near-IR data.  In this case, the high-resolution image is first convolved to the lower resolution seeing.  The
pixel-by-pixel flux of the ``blurred'' objects identified in the high-resolution image are then compared
directly to the pixel flux in the low-resolution image.  A minimization scheme determines the flux ratio between the two bands for every
high-resolution source, fitting all neighbors simultaneously to account for overlapping profiles in the low-resolution image.
This technique forms the basis of the {\sc TFIT} package \citep{laidler07} and the {\sc ConvPhot} software \citep{de-santis07} and
has been used widely \citep[e.g.,][]{papovich01, labbe05, grazian06}.  It has the advantage of making use of the available spatial
information, but does assume---as all methods do to some extent---that the shape of the flux profiles are independent of
wavelength.  In this case, the shape is determined by the pixel-to-pixel flux in the single, high-resolution band.  It can thus be
subject to noise spikes and other anomalies.

The final method is ``profile fitting'' which generalizes the ``flux fitting'' defined above.  The most widely-used example is the
photometry from SDSS, specifically the ModelMag photometry which is described in \citet{strauss02}, with more information
available on the internet\footnote{See {\tt http://www.sdss.org/dr7/algorithms/photometry.html} and {\tt
    http://www.astro.Princeton.EDU/\~rhl/photo-lite.pdf}}.  Here a smooth intrinsic profile is determined for every source,
usually in the deepest band.  In each of the other bands, this profile is convolved with the local PSF and scaled to the pixel
flux in that band.  Assuming the profile is wavelength-independent and correct, this method maximizes the signal-to-noise (S/N) on
the flux measurement and obtained colors.  But the adopted profile need not be a perfect match to the observed profile.  In this
case it can be thought of as a spatial weighting function applied to the total flux.  This is the approach taken by \citet{kuijken08} for measuring the
``Gaussian-aperture-and-PSF'' (GaaP) flux, a seeing-independent flux estimator.  Here the profile fitting is done to the shapelet
basis function with fits obtained in every band.  The shapelets can then be convolved analytically to a common PSF under which the
GaaP flux and resulting colors are defined.  Since the shapelet fitting is performed in every band, however, the photometric
scatter should be larger than in methods that fit all fluxes to a single intrinsic profile.

Each of these methods has strengths and weaknesses, but all of them suffer from a major problem when considering the size and pace
at which new imaging data is becoming available.  Multiband imaging data over thousands of square degrees can quickly sum to
$\sim$10 terabytes or more.  If new data are continuously being added, keeping the matched photometry up-to-date can be extremely
expensive computationally.  In addition to the raw computing time, the human cost of mastering and manipulating so much raw data
is significant.  Instead, it would be valuable to have the option of measuring the photometric properties of sources in a given survey and filter one time
only, but in a way that allows for a straightforward comparison with other filters and surveys without revisiting the pixel data.

In this work, we present an approach with this goal in mind that delivers ``synthetic aperture magnitudes'' ({\synmag}s), enabling
very fast matched photometry with reasonable precision.  Higher precision results can be obtained with a full analysis of all
available pixel data (for example, using profile fitting methods), but {\synmag}s provide a fast alternative until such an
analysis can be undertaken.  Our method works entirely at the catalog level.  We require recorded parameters for profile fitting
in at least one band, with measured aperture magnitudes (of arbitrary size) and PSFs recorded for the other bands.  As a test
case, we derive matched photometry from the combination of SDSS and the UKIRT Infrared Deep Sky Survey's Large Area Survey
(UKIDSS/LAS) and show that its performance is as good, if not better, than PSF-homogenized methods.  The \synmag\ technique is
easily applied to any data set that overlaps with SDSS, but is applicable generally to surveys employing profile-fitting
photometry in at least one band, such as CS82 (Erben et al., in preparation), HSC, and DES.  We make the \synmag\ code publicly
available.  In a followup paper (Higgs et al., in preparation), we present a second methodology specifically tuned to the
SDSS-UKIDSS comparison that offers additional flexibility.

An overview of the {\synmag} method is presented in Section \ref{method}, with specific notes relevant to SDSS given in Section
\ref{sdss}.  The use of Multi-Gaussian Expansions is described in Section \ref{mge}, while Section \ref{tests} presents various
tests of {\synmag}s applied to real data.  In Section \ref{discussion} we conclude with some discussion of ``application-oriented
photometry'' and describe how future surveys can provide catalog information to facilitate the use of {\synmag}s.  Appendix \ref{cookbook}
provides recipes for applying the {\synmag} software to real data.

\section{Method}\label{method}

\subsection{Overview}\label{method:overview}

We begin with an overview of our approach.  While our specific motivation is to apply this method to obtain matched photometry
between SDSS and UKIDSS imaging data sets, the methodology will be appropriate for many other applications.  As in nearly all
treatments of multiband photometry including the use of matched-aperture photometry, we will assume there are no color gradients.
This is obviously a poor assumption because it ignores how color depends on aperture size and obscures the meaning of an
integrated color (see Section \ref{discussion}), but is standard practice and simplifies the problem.

Consider two imaging data sets that we will refer to as $D_A$ and $D_B$.  They account for two sets of filter bands, obtained in
different conditions with different PSFs, perhaps on different telescopes.  In the specific case we describe later, $D_A$ will
refer to the SDSS DR7 data set and $D_B$ to UKIDSS.  Our method requires that profile fitting has been performed on the imaging
data from at least one of the filter bands in one of the data sets.  We will assume model profile fit parameters are available in
the $D_A$ catalog and will choose a definitive band for these profiles, $p$, which is typically the deepest band in $D_A$.  In
SDSS, $p$ is defined as the $r$-band.  Over all filter bands, $i$, available from both data sets, our goal is to obtain a reliable
set of matched colors, $C_{pi}$, where $C_{pi} = m_p - m_i$.

We assume profile fits are not available in $D_B$, otherwise matched colors could be obtained by comparing the profile fits in
$D_A$ to $D_B$.  Instead, we assume that only aperture photometry has been measured and recorded in the object catalogs for $D_B$.
This is often the case because aperture photometry is far easier to perform than profile fitting.

For a given band, $i$, we require a catalog measurement from $D_B$ of the spatially-dependent PSF, or PSF$_i$.  We then convolve
the fitted 2D profile $\Sigma_p$ from $D_A$ with PSF$_i$ and integrate the result from $R=0$ to $R=R_{\rm aper}$, where $R_{\rm
  aper}$ is the radius used to define the aperture photometry recorded in $D_B$.  Note that we have {\em not} assumed circular
symmetry.  While the apertures used are circular\footnote{In practice, our method could be extended to work with non-circular
  apertures.}, our method works with 2D profile fits that describe the varied, projected axis ratios of galaxies on the sky.
We do assume that the true flux is well described by smoothly declining profiles, however.  In this way, we have ``synthesized'' the
aperture flux that would have been measured in the $p$-band (from $D_A$) under the same seeing conditions as the $i$-band (from
$D_B$) and within the same aperture as used to measure aperture photometry in the $D_B$ catalog.  We use $m_{p,\rm SYN,i}$ to
refer to the resulting synthetic magnitude or ``{\sc synmag}.''  We can then construct the full set of colors across $D_A$ and
$D_B$ as $C_{pi} = m_{p,\rm SYN,i} - m_{i,\rm aper}$.

In some cases, as in SDSS, colors internal to $D_A$ or $D_B$ may already be available and more reliable than the {\sc synmag}
colors.  These can be substituted for the {\sc synmag} colors when available because the assumption of no color gradients means
that, given two filter bands, a galaxy by definition has only one color.  In practice, of course, {\sc synmag} colors will differ
from other techniques (e.g., ModelMag colors), but we emphasize that our goal is not to obtain the {\em best} set of matched
colors, but the optimal one given constraints on both computational time and the difficulty of downloading, understanding, and
organizing large imaging data sets.  Taking advantage of pre-computed colors internal to one of the data sets can lead to an
overall improvement in the characterization of a galaxy's true colors.  It is also possible to obtain colors from profile-fitting
magnitudes derived in different bands and even with different profile shapes by comparing the flux integral of the fit profiles at
a radius that encloses most of the light.  If the assumed profile shapes are significantly different, however, biases may be
introduced\footnote{A sense of the magnitude of such biases is given by the difference between Exponential and de Vaucouleurs
  magnitudes in SDSS, roughly 0.2 magnitudes $\pm$ 0.1 for galaxies near $r = 20$.} that depend on galaxy morphology and
concentration.

We can also use the set of colors, $C_{pi}$, to generate a set of PSF-matched magnitudes.  Let us assume that well-defined ``total'' magnitudes,
$m_{p,\rm tot}$, are available for the $p$-band in $D_A$, perhaps again via profile fitting.  In SDSS, these could be defined as
$r_{\rm ModelMag}$ or $r_{\rm CModelMag}$.  We can then define a new set of total-magnitude, PSF-matched photometry for each band, $i$, as
$m_{i,\rm tot} = m_{p,\rm tot} - C_{pi}$.  

Finally, we note that in the typical $D_B$ data set, aperture photometry is performed in several apertures of varying size.  It is
possible to compute {\sc synmag}s and resulting colors for each of these apertures.  A weighted average could produce a better
color measurement, and in the limit of very many apertures, this is equivalent to fitting the intrinsic profile, $\Sigma_p$, to
the circularly-averaged radial flux profile in the band $i$.  This moves closer to the construction of ModelMags in SDSS, in which
$\Sigma_p$ is fitted to the 2D pixel data in band $i$.  We do not pursue this extension here, however, in part because aperture
photometry spanning only a few pixels near the centers of objects requires a careful treatment of the flux variation {\em within}
pixels.  Even with well-resolved sources, these central aperture fluxes may not be computed reliably by many survey pipelines.
Because they represent regions of high surface brightness, this may bias attempts to fit $\Sigma_p$ directly to the tabulated
aperture flux measurements.

\subsection{Application to SDSS and UKIDSS}\label{sdss}

Synthetic aperture photometry is valuable in a variety of contexts, especially as an effective way to derive
a first set of matched photometry ahead of a more detailed analysis involving the pixel data.  Our original motivation, however,
came from the desire to obtain optical through near-IR colors for galaxies targeted by the Baryon Oscillation Spectroscopic Survey
(BOSS).  We desired matched colors between the SDSS imaging and UKIDSS.  In this section we provide relevant details about these
specific datasets, especially SDSS.  These are not only important for understanding the ways we have tested our approach, they
will be useful for applying synthetic aperture photometry to the overlap between SDSS and other data sets.

The SDSS imaging is reduced and processed by the {\sc Photo} pipeline, which has been packaged in {\tt sdss3tools}.  The source
detection, flagging, profile fitting, and photometry performed by {\sc Photo} is described on the SDSS website\footnote{For DR7,
  see {\tt http://www.sdss.org/dr7/algorithms/index.html}} under ``Algorithms.''  The SDSS camera is described in \citet{gunn98},
the telescope system in \citet{gunn06}, and the filter system in \citet{fukugita96}.  A technical summary is provided in
\citet{york00}.  Most relevant for our purposes is the profile fitting and definition of ModelMags, which are the default
recommendation for obtaining extended-source colors within SDSS.

A PSF-convolved, 2-dimensional de Vaucouleurs (deV), exponential (Exp), and PSF profile is fit to every detected source in every
band, $ugriz$.  For the deV and Exp profiles, the best-fit effective radius (along the major axis), axis ratio and position angle,
and normalization in the form of a total magnitude---as well as the estimated errors on these quantities---are recorded in the {\sc
  PhotObj} tables in the SDSS database.  The standard form for the exponential profile, with $r$ measured along the major axis, is given by

\begin{equation}
I_{\rm Exp}(r) = I_{e,\rm Exp} \exp\left[-\kappa_{\rm E} \left({\displaystyle \frac{r}{R_e}}-1 \right) \right],
\end{equation}

\noindent where $\kappa_{\rm E} = 1.67835$, as needed to set the definition of $\Sigma_e$ to the surface
brightness at the half-light or effective radius, $R_e$.  In {\sc Photo} v5\_6 the Exp profile is truncated at large radii in the
following manner:

\begin{equation}\label{eqn:Exp-SDSS}
I_{\rm Exp}^{\rm SDSS}(r)= \left\{ 
\begin{array}{l l}
  I_{\rm Exp} & \quad \mbox{$r/R_e < 3$}\\
  I_{\rm Exp} [1 - (r/R_e - 3)^2]^2 & \quad \mbox{$3 < r/R_e < 4$}\\
  0 & \quad \mbox{$r/R_e > 4$}\\ \end{array} \right.
\end{equation}

For the deV profile, ``softening'' is applied in the center so that the profile is given as

\begin{equation}
I_{\rm deV}(r) = I_{e,\rm deV} \exp\left[-\kappa_{\rm d} \left( \left[ \left({\displaystyle \frac{r}{R_e}}\right)^2 + 0.0004
    \right]^{\frac{1}{8}} -1 \right)\right],
\end{equation}

\noindent with $\kappa_{\rm d} = 7.66925$.  The truncation for the deV profile is accomplished as follows:

\begin{equation}\label{eqn:deV-SDSS}
I_{\rm deV}^{\rm SDSS}(r)= \left\{ 
\begin{array}{l l}
  I_{\rm deV} & \quad \mbox{$r/R_e < 7$}\\
  I_{\rm deV} [1 - (r/R_e - 7)^2]^2 & \quad \mbox{$7 < r/R_e < 8$}\\
  0 & \quad \mbox{$r/R_e > 8$}\\ \end{array} \right.
\end{equation}

\noindent These definitions can be found in {\tt makeprof.c} in the {\sc Photo} package.  We note that some examples of specific
SDSS profiles are contained in {\tt phFitobj.h}, but the values given there for the total flux after accounting for the imposed
truncation can only be reconciled if the truncation follows the form $[1 - (r/R_e - 3)^2]^{1.0}$ for Exp and $[1 - (r/R_e - 7)^2]^{1.0}$
for deV (that is, the terms are not squared).  We instead take as correct the forms hard-coded in {\tt makeprof.c} and described by Equations \ref{eqn:Exp-SDSS} and
\ref{eqn:deV-SDSS}.  In terms of total magnitudes, the truncations lead to differences of 0.01--0.03 mag compared to the standard
forms.  Meanwhile, the applied profile softening at the center of the deV profile is more significant compared to the standard profile and can lead to
differences of 0.05 mag.

The deepest band of imaging in SDSS is the $r$-band which is used to determine the profile-matched photometry known as ModelMags.
These are defined as follows.  For each object, the better of the $r$-band Exp or deV fits as measured in all the filters is taken as the
default profile for that object.  This profile is then scaled in normalization (but in no other parameters) to the pixel data in
the other bands at the source location.  The integral of the total flux of the resulting best fit in each band defines the
ModelMag in that band.  ModelMag photometry provides galaxy colors with less scatter than other estimators, a strong motivation
for the {\synmag} approach which enables the use of ModelMags in conjunction with photometry matched to other imaging data sets.

\subsection{Mock Image Implementation}\label{mock}

In the next two sections we outline two methods for implementing synthetic aperture photometry.  The first
approach is more ``brute force'' in nature but provides a simple way of testing the more sophisticated Multi-Gaussian Expansion
implementation that follows.  

For each source, we take the profile fit parameters describing the intrinsic profile, $\Sigma_p$, from $D_A$ catalog.  These parameters
should include a characterization of the profile shape (e.g., exponential, de Vaucouleurs, or more generally a \sersic-$n$ value),
the effective radius, $R_e$, the profile normalization (e.g., the surface brightness, $\Sigma_e$, at $R_e$) and ideally the axis ratio.
We then use these parameters to create a mock postage stamp image of $\Sigma_p$ with zero noise.  The mock image is convolved with
the spatially-dependent PSF of the target band in $D_B$ and the flux in the now convolved profile is summed over the relevant
aperture(s) defined in $D_B$, returning the {\synmag}s in the $p$-band for that source, $m_{p,\rm SYN,i}$.

There are a number of obvious limitations to this implementation.  First, our goal has been to achieve a very fast way of
producing matched photometry.  The image convolution here is the slowest step, and while we have avoided returning to the raw
pixel data, the mock image implementation is not that different than working with the real data since both require
convolutions performed on 2D images.  It is not clear that the ease of this approach justifies giving up the advantages of a
full-fledged re-analysis of the imaging in $D_A$ and $D_B$.

A second problem compounds the issue.  Because our starting point is the intrinsic profile, $\Sigma_p$, the pixel scale
``resolution'' of the mock image has to be high enough to accommodate the steeply rising interiors of profiles with small $R_e$,
especially deV profiles or those with high \sersic-$n$ values.  In our brute force approach, we found the needed resolution simply
by decreasing the pixel scale of the mock images until there were no differences in the {\synmag}s measured at different
resolutions.  In tests of massive galaxy profiles at $z \sim 0.5$ (see Section \ref{tests:fibermags}) we found convergence at a
mock image resolution of roughly 0.01 arcsec per pixel.  This resolution results in prohibitively large PSF convolution times that
were the whole motivation of synthetic photometry.  A more clever approach with an adaptive resolution grid would address this
problem, but given the other limitations, we adopt a completely different and more elegant approach in the following section.

\subsection{Multi-Gaussian Expansion Implementation}\label{mge}

Synthetic aperture photometry becomes attractive compared to a reanalysis of imaging pixel data once it is significantly faster.
Because the slowest step in methods that work with pixel data is often PSF convolution, we develop a method here which decomposes
intrinsic profiles and PSFs into multiple Gaussian expansions (MGEs).  The fact that convolutions between Gaussians can be
calculated analytically makes synthetic photometry with MGEs extremely fast, as long as the MGE decomposition of the profile and
PSF need only be performed once for the full data set.  We will show this is commonly the case.

The power of MGEs in the context of smoothly decreasing profiles common in astronomy was recognized early on
\citep[e.g.,][]{trumpler53} and applied rigorously to flux profiles such as the King model and Moffat PSF by \citet{bendinelli91}.
Their utility and a modern implementation was presented in \citet{emsellem94} and also discussed in \citet{connolly00}.

\subsubsection{MGE Decomposition}\label{decomposition}

\begin{figure*}
\plotone{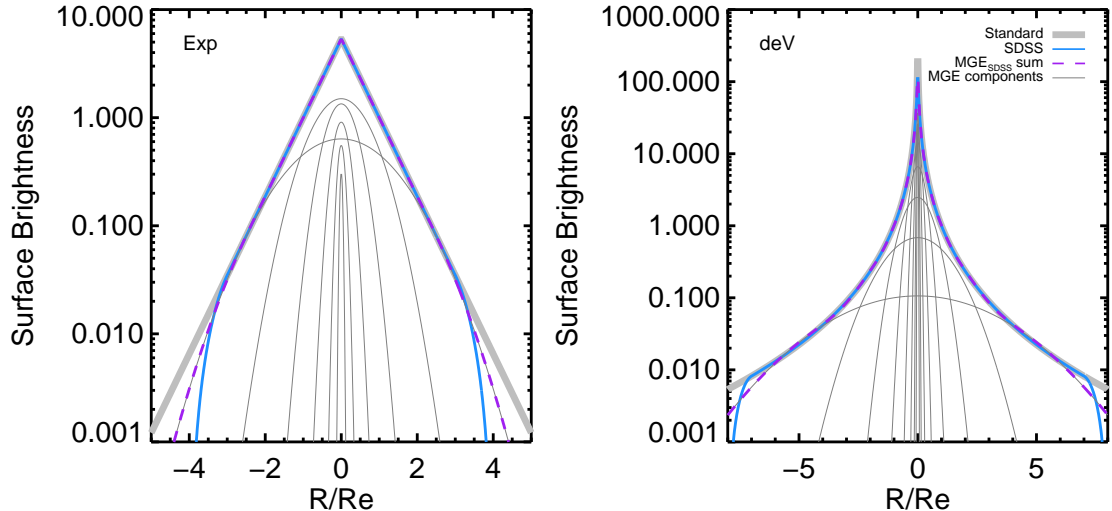}
\caption{A sample Exponential (left) and de Vaucouleurs (right) profile, illustrating the MGE decomposition.  The standard forms
  are shown as solid purple lines, while the SDSS-modified forms are shown in dark grey.  Surface brightness units are arbitrary
  and scaled so that $\Sigma(R_e) \equiv \Sigma_e = 1.0$.  The MGE components for the modified
  profiles are overplotted as thin, light grey lines.  They sum to give the dashed blue curves.
  \label{fig:mge}}
\end{figure*}

Our first step is to describe the intrinsic profiles, $\Sigma_p$, that are fit to the profile-band data in the dataset,
$D_A$, by an MGE, that is, a sum of $j$ Gaussian profiles:

\begin{equation}
\Sigma_p(r) = \sum_j \Sigma_{G,j}(r) = \sum_j A_j \exp\left(-\frac{r^2}{2\sigma_j^2}\right)
\end{equation}

\noindent where $A_j$ is the normalization (central flux density) and the width of each Gaussian is given by $\sigma_j$.  The
summed flux of each Gaussian is $L_j = 2\pi A_j\sigma_j^2$.  As we describe below, it is possible to determine an MGE for a profile of
arbitrary size and shape, a process that takes less than a minute.  However, as long as the chosen intrinsic form of the profile
scales with an effective radius ($R_e$) and normalization ($\Sigma_e$), the $\sigma_j$ and $A_{j}$ of a single MGE determined for that
profile form can also be scaled in the same way.  For a defined profile shape, this means a single MGE can be scaled to match
profiles with arbitrary size and flux.  Determining the MGE for the intrinsic profile of every source in $D_A$ is not required,
although a separate MGE is needed for each profile {\em shape} that was fit for.  In SDSS, this means an MGE must be determined
for both the Exp and deV profiles.  For {\sersic} profiles generally, an MGE is required for each \sersic-$n$ value.  From the
point of view of {\synmag}s, it is helpful if the number of profile shapes fitted for in $D_A$ (i.e., the number \sersic-$n$
values fitted for) is restricted to a modest number.  The MGE descriptions of these shapes can be determined once and then used
for any subsequent measurement of {\synmag}s involving the datasets in question.

So far we have considered axisymmetric profiles defined in one dimension.  Two dimensional profiles are easily described by
a sum of 2D Gaussians.  We let the profile's major axis be the x-axis and define the axis ratio, $q = b/a$, where $a$ and $b$ are
characteristic sizes along the major and minor axes respectively.  The MGE becomes,


\begin{equation}\label{eqn:mge}
\Sigma_p(x,y) = \sum_j A_j \exp\left(-\frac{x^2}{2\sigma_{x,j}^2} -\frac{y^2}{2q^2\sigma_{x,j}^2} \right).
\end{equation}

\noindent after substituting $\sigma_y = q \sigma_x$.  For a given profile shape, any 2D representation of arbitrary size, flux, or axis ratio can thus be described by a
single 1D MGE.

The MGE must now be carefully determined with an eye towards minimizing the number of Gaussian terms while preserving precision in
the MGE approximation of the intrinsic profile.  We find that a least-squares fit to the full (circular) 2D profile provides the
best approximation to the true surface brightness profile because the fit is then weighted by the intensity.  MGE approximations
for standard Exp and deV profiles as well as the modified profiles used in SDSS are given in Table \ref{table:mge} and their 1D
radial profiles are shown in Figure \ref{fig:mge}.  In terms of enclosed flux, the number of terms used in these MGEs provide
accuracies at the 0.01 mag level out to $10 R_e$ and so we deem them adequate approximations for our purposes.  More details on
the optimal determination of these fits as well as more accurate representations employing additional Gaussian terms are provided
in \citet{hogg12}.

It is generally difficult to fit the imposed truncation in the modified SDSS profiles with Gaussians, although the central smoothing in the
modified de Vaucouleurs profiles makes an MGE solution easier in the central regions.  The MGEs begin to deviate from the true SDSS
profiles at radii larger than roughly $3.5 R_e$ for the Exp profile and $6 R_e$ for the Dev profile (see Figure \ref{fig:mge}).
This introduces aperture flux discrepancies of roughly 0.01--0.02 mag, and so we consider these MGEs adequate for our purposes.

\begin{deluxetable}{lcccccccc}
\tablecaption{MGE Parameters}
\tabletypesize{\footnotesize}
\tablewidth{0pt}
\tablecolumns{9}
\tablehead{
\multicolumn{1}{c}{} & \multicolumn{2}{c}{$\Sigma_{\rm Exp}$} & \multicolumn{2}{c}{$\Sigma_{\rm deV}$} & \multicolumn{2}{c}{$\Sigma_{\rm Exp}^{\rm SDSS}$} & \multicolumn{2}{c}{$\Sigma_{\rm deV}^{\rm SDSS}$} \\
\cline{2-3} \cline{4-5}  \cline{6-7}  \cline{8-9} \\
\colhead{N$_{\rm terms}$} & \colhead{$\sigma_j$} & \colhead{$A_{j}$} & \colhead{$\sigma_j$} & \colhead{$A_{j}$} & \colhead{$\sigma_j$} & \colhead{$A_{j}$} & \colhead{$\sigma_j$} & \colhead{$A_{j}$} \\
}

\startdata
1  &  5.1  &  0.45  &  0.1  &  328.94  &  3.5  &  0.31  &  1.5  &  30.33  \\
2  &  13.8  &  0.80  &  0.5  &  176.20  &  9.4  &  0.55  &  3.2  &  38.13  \\
3  &  28.8  &  1.22  &  1.5  &  99.93  &  19.8  &  0.91  &  6.5  &  26.07  \\
4  &  53.2  &  1.46  &  3.9  &  52.88  &  37.4  &  1.34  &  13.0  &  14.27  \\
5  &  91.2  &  1.04  &  9.9  &  24.55  &  67.9  &  1.50  &  26.2  &  6.59  \\
6  &  150.2  &  0.22  &  24.7  &  9.29  &  122.5  &  0.64  &  53.6  &  2.47  \\
7  &  --  &  --  &  63.9  &  2.51  &  --  &  --  &  115.5  &  0.68  \\
8  &  --  &  --  &  192.6  &  0.35  &  --  &  --  &  289.9  &  0.11  \\

\enddata
\label{table:mge}
\tablecomments{The components are normalized so that the sum yields $\Sigma(R_e) \equiv \Sigma_e = 1.0$ in arbitrary units of surface flux density.  The units
  of $\sigma_j$ are $0.01 R_e$.}
\end{deluxetable}

\subsubsection{PSF Convolution with MGEs}

The {\em observed} galaxy profile, $I_p$ is the convolution of $\Sigma_p$ with the PSF profile, $R_{\rm PSF}$.  The Fourier transform, $\mathcal{F}$, of this
convolution can be expressed by the product of Fourier transforms:

\begin{equation}\label{eqn:convolution}
\mathcal{F}(\Sigma_p \otimes R_{\rm PSF}) = \mathcal{F}(\Sigma_p) \mathcal{F}(R_{\rm PSF})
\end{equation}

\noindent The power of the MGE approach lies in the fact that the Fourier transform of a Gaussian is also a Gaussian and
easily determined analytically.  Fourier transforms are also linear, so that for each Gaussian term, $j$,

\begin{equation}
\mathcal{F}(\Sigma_p(r)) = \sum_j \int A_j \exp \left(-\frac{r^2}{2\sigma^2_j} - 2\pi i \vec{k_r}\right) {\rm d}r.
\end{equation}

Referring to Equation \ref{eqn:mge}, $\mathcal{F}(\Sigma_p(x,y))$ can be expressed as

\begin{equation}
\mathcal{F}(\Sigma_p(x,y)) = 2\pi \sum_j A_j \sigma_j^2 q \exp \left(-\frac{2\pi^2 \sigma_{x,j}^2}{2}\left(k_x^2 + q^2 k_y^2 \right) \right)
\end{equation}

\noindent For the derivations here, we assume that $R_{\rm PSF}$ can also be decomposed into a series of circular Gaussian terms, $k$, with coefficients
$R_k$ and standard deviations, $\sigma_{R,k}$.  A 2-component Gaussian PSF, often used to account for power in the wings of the
PSF, is an obvious example.  Taking the inverse Fourier transform of Equation \ref{eqn:convolution}, we can express the
PSF-convolved ``observed'' profile, $I_p$, as

\begin{equation}\label{eqn:invft1}
\begin{split}
& I_p(x,y) = \\
& \sum_j \sum_k C_{j,k} \exp \left(-\frac{x^2}{2(\sigma_{x,j}^2 + \sigma_{R,k}^2)} -\frac{y^2}{2(q^2 \sigma_{x,j}^2 + \sigma_{R,k}^2)} \right),
\end{split}
\end{equation}

\noindent where the coefficients $C_{j,k}$ are given by

\begin{equation}\label{eqn:invft2}
C_{j,k} = \frac{2\pi A_j R_k q \sigma_{x,j}^2 \sigma_{R,k}^2} {\sqrt{(\sigma_{x,j}^2 + \sigma_{R,k}^2)(q^2 \sigma_{x,j}^2 + \sigma_{R,k}^2)}}.
\end{equation}

While we have assumed circular PSF components to this point, the {\synmag} approach and associated software handles non-circular, spatially offset, and even negative components, providing a flexible way to describe more complicated PSF shapes via $R_{\rm PSF}$.  

\subsubsection{MGE Aperture Flux and profile}\label{sec:mgestr}

Equation \ref{eqn:invft1} provides an analytic approximation, $I_p$, of a galaxy light profile with an intrinsic shape measured in
the $p$-band, but observed with the PSF of another band.  In our case, the second band comes from a different dataset, $D_B$ for
which only aperture photometry is available.  To compare the photometry recorded in $D_B$ to the aperture flux, $F_p$, that would have
been measured in the $p$-band with the same PSF, we must integrate Equation \ref{eqn:invft1} over the aperture(s) used in $D_B$.
Because $I_p$ is generally elliptical while the apertures are circular, a numerical
integration is required.  Working in Cartesian coordinates, the 2D integral of Equation \ref{eqn:invft1} over a circular aperture
of radius $R$ can be expressed as

\begin{equation}
\begin{split}
& F_p(R) = \\
& \sum_j \sum_k \int_{0}^R \!\!\! \int_0^{c(x)} 4 C_{j,k} \exp\left(-\frac{x^2}{2\hat{\sigma}_x^2} \right)\exp\left(-\frac{y^2}{2\hat{\sigma}_y^2} \right) {\rm d}x{\rm d}y,
\end{split}
\end{equation}

\noindent where $c(x) = \sqrt{R^2-x^2}$, and $\hat{\sigma}$ represents the effective standard deviation of a PSF-convolved MGE term, such that $\hat{\sigma}_x^2
= (\sigma_{x,j}^2 + \sigma_{R,k}^2)$, and $\hat{\sigma}_y^2 = (q^2 \sigma_{x,j}^2 + \sigma_{R,k}^2)$.  The second integral can be expressed in terms of the
error function, so that a fast 1D implementation of the numerical integral can be written as

\begin{equation}\label{eqn:flux_integral}
\begin{split}
& F_p(R) = \\
& 4  C_{j,k} \hat{\sigma}_y \sqrt{\frac{\pi}{2}} \int_{0}^R \exp\left(-\frac{x^2}{2\hat{\sigma}_x^2} \right){\rm
  erf}\left(\sqrt{\frac{R^2-x^2}{2\hat{\sigma}_y^2}}\right) {\rm d}x.
\end{split}
\end{equation}

We can also use the MGE description of $I_p$ to estimate the effective radius and axis ratio of the {\em observed} flux profile.  Working along the major
axis---or equivalently assuming the axis ratio is 1.0---we can define an effective half-light radius, $R_{e,{\rm maj}}$, analogous to the $R_e$ defined for
the intrinsic profile, $\Sigma_p$.  The integral of the 2D flux is easily calculated:

\begin{equation}
L(x) = \sum_{j,k} 2\pi C_{j,k} \hat{\sigma}_x^2 \left[ 1 - \exp\left(\frac{x^2}{2 \hat{\sigma}_x^2}\right) \right].
\end{equation}

\noindent Setting this equal to the (major-axis) ``half-flux'' given by $L_{1/2, {\rm maj}} = \sum \pi C_{j,k} \hat{\sigma}_x^2$ and solving for $x$, one obtains
$R_{e,{\rm maj}}$.  In practice, this definition of half-light radius is not as intuitive as in the case of $\Sigma_p$ because the axis ratio depends
on radius (as we show below).  Thus, the minor-axis half-light radius is not simply related to $R_{e,{\rm maj}}$.  We can alternatively define a second measure of the
half-light radius, $R_{e,{\rm circ}}$, that corresponds to the size of a circular aperture which contains half of the total flux.  Here, we integrate Equation
\ref{eqn:flux_integral} until the enclosed 2D flux is equal to the true half-flux as given by $L_{1/2} = \sum \pi C_{j,k} \hat{\sigma}_x\hat{\sigma}_y$.

\begin{figure*}
\epsscale{1.15}
\plotone{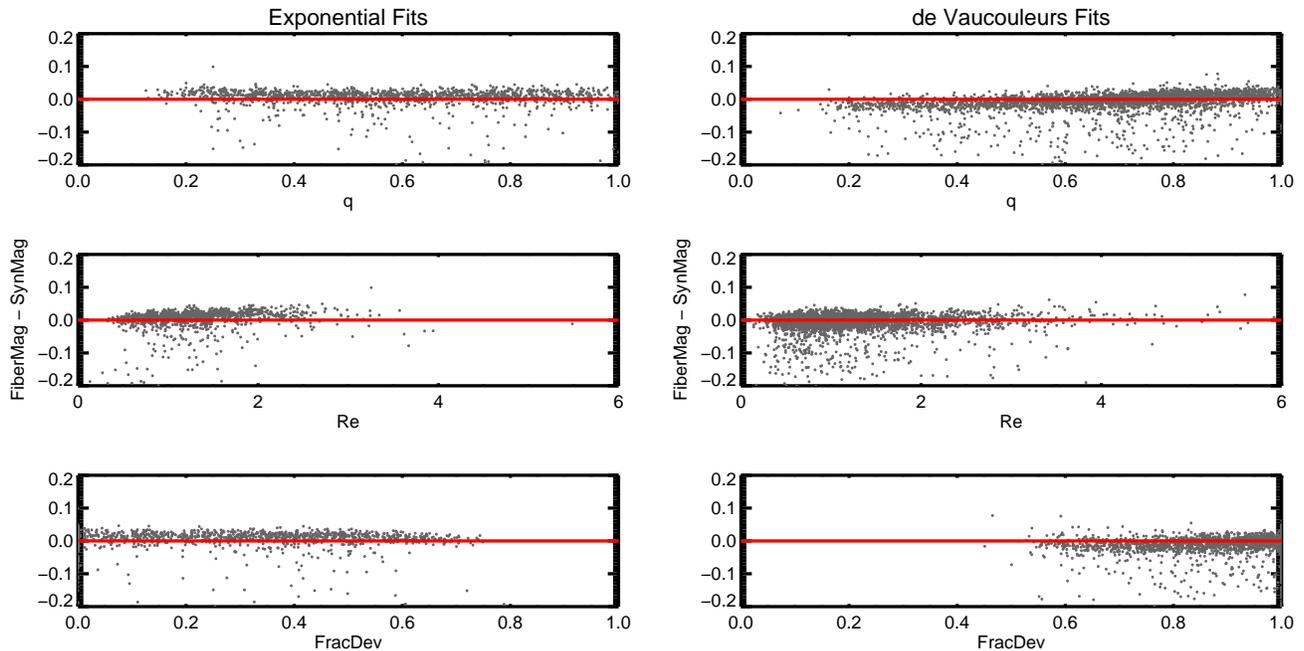}
\caption{Comparison between $r$-band FiberMags and {\synmag}s.  Sources best fit by exponential profiles are compared in the left-hand column, while de
  Vaucouleurs fits are shown in the right-hand column.  From top to bottom, the panels investigate magnitude differences as a function of intrinsic axis
  ratio ($q$), effective radius ($R_e$), and FracDev, an estimator for how well the profile shape can be fit by a de Vaucouleurs profile.  Residuals at the
  0.01--0.02 mag level are probably caused by an imperfect treatment of the PSF.
  \label{fig:fibermag}}
\end{figure*}

Turning to the effective axis ratio, while the intrinsic axis ratio is fixed, the observed value will be a function of radius.  Seeing will round out profiles at a scale
corresponding to the PSF FWHM, but will have a smaller effect on elongated (or inclined) profiles at larger radii.  The {\em convolved} axis ratio of each
Gaussian term is given by

\begin{equation}
\displaystyle \hat{q}_{j,k} = \frac{\hat{\sigma}_y^2}{\hat{\sigma}_x^2} = \frac{q^2 \left({\displaystyle \frac{\sigma_{x,j}^2}{\sigma_{R,k}^2}} \right) + 1}{\left({\displaystyle \frac{\sigma_{x,j}^2}{\sigma_{R,k}^2}} \right) + 1}.
\end{equation}

To derive an estimate of the observed profile's axis ratio, $\langle q \rangle$, we can weight each term by approximately the half-flux:

\begin{equation}
\langle q \rangle = \frac{\sum w_{j,k} \hat{q}_{j,k}}{\sum w_{j,k}},
\end{equation}

\noindent where the weights are given by $w_{j,k} = C_{j,k} \hat{\sigma}_x\hat{\sigma}_y$.  

\section{Tests with Galaxy Photometry}\label{tests}

The motivation for {\synmag}s is fast matched photometry for overlapping data sets that span large areas of the sky and thus make
a reanalysis of pixel data difficult and time consuming.  Since we cannot use the stellar locus---a common way of testing photometry methods---we
perform tests of the {\synmag} approach using galaxy magnitudes and colors.  We first show that synthetic magnitudes are
consistent with SDSS FiberMags.  We then compare galaxy colors from synthetic photometry to ``FWHM-homogenized'' matched-aperture
colors from the GAMA Survey.  We consider the derived scatter in the color of the red sequence and in the photometric redshifts
determined from these two sets of photometry, finding that {\synmag}s perform as well or better than the standard but more
laborious FWHM-homogenization method.  The latter two tests require redshift information which we take from the Baryon Oscillation
Spectroscopic Survey (BOSS).

\subsection{SDSS FiberMags}\label{tests:fibermags}

In all five bands, and for every detected source, the SDSS {\sc Photo} pipeline provides a ``FiberMag'' estimate of the 3\arcsec\ diameter
aperture magnitude, assuming seeing of 2\arcsec\ FWHM.  This is meant to approximate the flux measured by a spectroscopic fiber in SDSS-I/II.  Since {\synmag}s
can provide synthetic aperture photometry for an aperture of arbitrary size and with arbitrary seeing, a natural first test is to compare {\synmag}
approximations of FiberMags to their actual values.

\begin{figure*}
\plotone{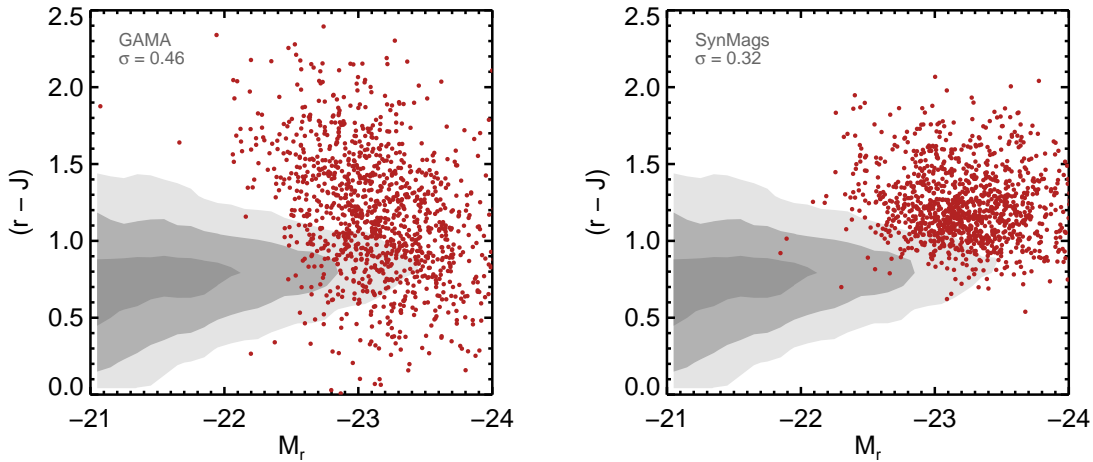}
\caption{Restframe $(r-J)$ color-magnitude diagrams of (mostly) red-sequence CMASS galaxies from the BOSS survey. The observed
  scatter for the GAMA FWHM-homogenized matched-aperture photometry is larger than for the {\synmag} approach described here.
  Contours indicate the location of much less luminous COSMOS galaxies in the same redshift range.
  \label{fig:redseq_rj}}
\end{figure*}

With an eye towards future applications of {\synmag} photometry for BOSS galaxies, we perform this exercise using a sub-region of the SDSS Stripe 82 Coadd catalog
\citep{annis11} with $355^{\circ} < {\rm RA} < 360^{\circ}$.  We select extended sources with non-stellar ``psfmag'' colors and detections in the UKIDSS LAS
catalog (DR4).  This effectively selects galaxies with $M_* > 10^{11} \msun$ and $0.2 < z < 0.7$.  We apply cuts on both SDSS and UKIDSS flags to ensure a clean
sample and remove blended objects by excluding SDSS sources with the {\sc blend} flag set.  The exact nature of the flag cuts do
not influence our results.

We have found that a precise understanding of the PSF is important for resolving discrepancies at the 0.05 mag level.  In the SDSS imaging, the PSF can be
approximated by a 2-component Gaussian.  The second component has a width that is twice that of the first component, but accounts for only 10\% of the
PSF power.  In the {\sc Photo} pipeline (see {\tt measureObj.c}), the FiberMag is determined by convolving a sub-image by a single Gaussian kernel with
$\sigma_k^2 = \sigma_{2.0}^2 - \sigma_{\rm mom}^2$, where $\sigma_{2.0}$ is the value of $\sigma$ for a Gaussian with ${\rm FWHM} = 2.0$\arcsec\ (i.e., $2.0/2.3548$)
and $\sigma_{\rm mom}$ is the adaptive second moment size of the local PSF (referred to as {\tt M\_rr\_cc}).  Thus, the FiberMags are {\em not} measured on images with a FWHM
2.0\arcsec\ Gaussian PSF but with a more complicated PSF profile with significant power in the wings.

For the consistency check here, we do not attempt to recover the applied kernel or the original (and likely spatially varying) PSF of the underlying image.
Instead, we approximate the final, effective PSF for FiberMags as a 2-component Gaussian (as described above), with the FWHM of the first term set to
2.05\arcsec.  Experimentation revealed that our uncertainty in this choice introduces systematic offsets at the 0.01-0.02 mag level.  We then
divide the sample into sources that are best fit by exponential profiles versus de Vaucouleurs profiles (this is easily determined by whether the
ModelMag matches the ExpMag or deVMag) to see how the {\synmag} MGEs and PSF convolutions perform in either case.  

The results are shown in Figure \ref{fig:fibermag} which compares $r$-band magnitude differences between {\synmag}s and FiberMags
as a function of axis ratio, size, and FracDev, a quantity that expresses how well a source can be described by a de Vaucouleurs
profile.  The {\synmag}s perform well with small systematic residuals at the 0.01--0.02 mag level that probably result from our
imperfect treatment of the PSF.  We expect additional systematics from the discrepancy between the MGE profile approximations and
the truncated SDSS profiles and from the fact that a single-component profile is not a good description of most galaxies.  For our
purposes, these systematics are less than the statistical uncertainties in most applications of galaxy photometry as well as the photometric
calibration uncertainties inherent in comparisons of SDSS to other data sets.  Finally, we note a small degree of scatter towards
brighter FiberMags.  A large fraction ($\sim$80\%) of these objects were removed by restricting to non-blended sources.  We therefore surmise
that the remainder are objects with nearby neighbors that contribute flux to the FiberMag aperture or galaxies with bright
sub-components that are not accounted for in the profile fitting.

\subsection{Red Sequence}\label{tests:red}

Having verified the ability of {\synmag}s to reproduce aperture photometry measured in the SDSS $r$-band, we now investigate how
well {\synmag}s provide matched photometry across differing data sets.  Our focus will be on regions of overlap between SDSS
imaging and the UKIDSS/LAS.  These data sets have been matched previously using standard FWHM-homogenization with re-measured
aperture photometry in two recent papers.  \citet{matsuoka10} carry out this exercise for 40 deg$^2$ in Stripe 82.  The GAMA
survey \citep{driver11} does so in three equatorial fields, totaling 144 deg$^2$ \citep{hill11}.  Because the GAMA
DR1 catalogs are larger and publicly available\footnote{{\tt http://www.gama-survey.org/dr1/YR1public.php}}, we will use them to test the
synthetic photometry method developed here.

As described in \citet{driver11} and further detailed by \citet{hill11}, the GAMA DR1 photometric catalog was constructed by first
downloading the reduced images in all 9 available bands---$ugriz$ from SDSS, and $YJHK$ from LAS---and convolving all imaging to a
common seeing FWHM of 2.0\arcsec.  We refer to this as ``FWHM-homogenization,'' and distinguish it from ``PSF-homogenization'' in
which the FWHM and detailed shape of the PSF are made identical in every band \citep[see][]{hildebrandt12}.  All images are
registered to a common astrometry, and source detection is performed in the $r$-band, which defines the Kron aperture that is applied to
the flux measurements in the other bands.  In the G09 and G15 fields, the catalog is limited to a Petrosian magnitude of $r_{\rm pet} <
19.4$, while in G12 the limit is $r_{\rm pet} < 19.8$.  Further details and various tests of the performance of GAMA photometry are
provided in \citet{hill11}.

\begin{figure*}
\plotone{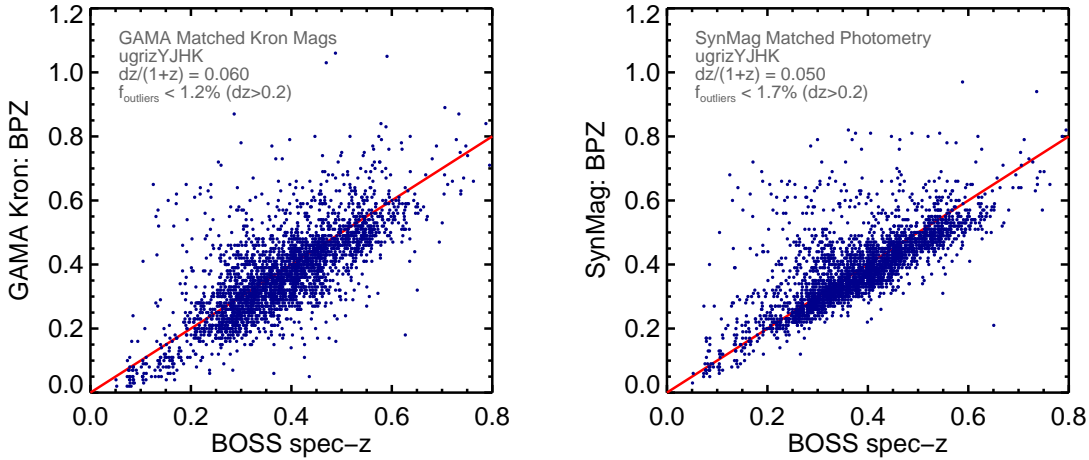}
\caption{For the same sample of BOSS galaxies, the quality of photometric redshifts derived from FWHM-homogenized $ugrizYJHK$ GAMA
  photometry (left panel) compared to the synthetic aperture photometry method using the same filter set (right panel).  The
  \synmag\ approach performs better.
  \label{fig:photoz_all}}
\end{figure*}

Our first test measures the scatter in the restframe $(r - J)$ colors of so-called CMASS galaxies from the Baryon Oscillation
Spectroscopic Survey \citep[BOSS, see][]{eisenstein11}, most of which lie on the red
sequence.  The BOSS target selection is briefly described in \citet{anderson12} and will be further detailed in Padmanabhan et al.~(in
preparation).  The CMASS selection is designed to target very luminous galaxies with $z > 0.4$ and $M_* \gtrsim 10^{11.4} \msun$.
Most have red colors, although 20--30\% show evidence for disk or irregular components, and 15\% have somewhat bluer
colors than a standard red-blue color cut \citep{masters11}.  We select 1203 CMASS galaxies in the three GAMA regions with secure
spectroscopic redshifts.  We apply K-corrections to infer restframe absolute magnitudes using {\tt K-correct} \citep{blanton07}
and apply no evolutionary corrections.

{\synmag}s are derived using the method described in previous sections from the SDSS DR7 \citep{abazajian09} and UKIDSS/LAS
publicly-available photometric catalogs\footnote{Special care is required in querying the UKIDSS database and removing corrections
  applied to the aperture photometry in the WFCAM Science Archive (WSA).  These will be described in detail in Bundy et al.~(in
  preparation).}.  Detections are required in all bands.  We chose to synthesize 2\farcs8 radius UKIDSS aperture photometry.
While we primarily use UKIDSS DR8 catalogs, the GAMA photometry was based on DR4, so we note that we find no differences when
using UKIDSS DR4 catalogs instead.  The {\synmag}s model the varying PSF in each of the $YJHK$ bands as a Gaussian with the local
FWHM taken from the LAS catalogs (note that a more sophisticated PSF treatment is possible).  The typical UKIDSS FWHM is 0\farcs8,
which is smaller than the 1\farcs1 seeing SDSS imaging that defines the intrinsic profiles needed for {\synmag} photometry.
Extinction corrections from \citet{schlegel98} are applied using values recorded in both the SDSS and UKIDSS catalogs.  Note that
the Kron-aperture-matched photometry from GAMA DR1 is also extinction corrected.

The restframe $(r-J)$ color-magnitude diagrams for both the GAMA photometry (left panel) and {\synmag} photometry (right panel)
are shown in Figure \ref{fig:redseq_rj}.  The choice of $(r - J)$ provides a test of the photometric matching between the SDSS and
UKIDSS data sets.  To provide a reference, the contours indicate the location of much less luminous COSMOS galaxies\footnote{The
  COSMOS data set is described in \citet{bundy10}, with restframe colors calculated by \citet{ilbert10}.} with similar redshifts
as the CMASS sample.  We note an offset that may arise from the combination of zeropoint calibration uncertainties in the COSMOS
data set \citep[see][]{capak07a} and the fact that the very massive BOSS galaxies are not well represented in the small volume of
the COSMOS field.  Because photometric scatter is likely to dominate over intrinsic scatter, the observed scatter in these
diagrams gives some indication of the performance of the photometry matching.  The {\synmag} photometry appears to do very well,
providing a smaller scatter ($\sigma = 0.32$) compared to the GAMA photometry ($\sigma = 0.46$).  We note however that GAMA DR1
photometry may have been affected by a bug in the PSF convolution (S.~Driver, private communication).  Our own tests suggest its
impact on the GAMA colors may not be significant, but future releases of GAMA photometry may show improvement.

\subsection{Photometric Redshifts}\label{tests:photoz}

The moderate scatter in the {\synmag} SDSS-UKIDSS colors of galaxies on the red sequence suggests that synthetic aperture
photometry performs well, but the scatter alone does not tell us whether the colors are ``correct.''  We therefore perform a
second test by comparing the performance of photometric redshifts derived using both the GAMA and {\synmag} approach.  Here the
goal of accurate \photozs\ is more clearly defined, although optimizing photometry based on the performance of \photozs\
carries implicit assumptions that should be carefully considered (Section \ref{discussion}).  

We again turn to BOSS galaxies with spectroscopic redshifts as a benchmark for evaluating the \photoz\ performance.  We utilize
all BOSS targets in this case (including the LOWZ sample) which allows us to probe redshifts from 0.1 to 0.8, thereby testing the
ability of multiple SDSS-UKIDSS filter combinations to correctly measure restframe spectral features. We apply the v1.99.3
Bayesian Photometric Redshift (BPZ) code \citep{benitez00,coe06} to both the GAMA matched photometry and the \synmag\ photometry.
BPZ includes a procedure for estimating photometric zeropoint offsets from the mismatch between the best-fit templates and the
observed photometry of galaxies where the redshift is known spectroscopically.  The resulting suggested offsets were $\sim$0.02
mag in the $u$ through $J$ bands, but 0.1 mag in the $H$-band, and 0.18 mag in the $K$-band, reflecting the poorer quality of
the BPZ templates in the near-IR.  We apply these offsets to both sets of photometry, although their impact is small.

The comparison of derived photometric redshifts compared to the 2989 BOSS-measured spectroscopic redshifts is shown in Figure
\ref{fig:photoz_all}.  We emphasize that the goal is not to obtain the best \photozs\ possible for this sample, but to
compare the resulting \photoz\ quality from two different photometry techniques.  When excluding outliers defined as
$|z_{\rm spec} - z_{\rm phot}| > 0.2$, we find that the FWHM-homogenization approach used by GAMA achieves a ${\rm dz}/(1+z)$ of
0.06 while the {\synmag}s deliver an improved scatter of 0.05.  Indeed, the decreased scatter in the \synmag\ photo-$z$s is
evident by eye in Figure \ref{fig:photoz_all}.  The {\synmag}s outlier fraction is slightly larger, however (1.7\% instead of
1.2\%).  

This result was anticipated by Figure \ref{fig:redseq_rj}; the tighter colors do appear to yield better \photozs.  This is highly
encouraging for the utility of the {\synmag} approach.  Even if future improvements to the GAMA photometry deliver better BPZ
performance, the {\synmag}s require far less work and computation time.  The photometry computation for the comparison set of
$\sim$3000 galaxies here takes less than one minute on a modern desktop (8-core CPUs with 10 GB memory).  Combined with the
red-sequence test, these results indicate that the scatter in {\synmag} colors may outperform PSF homogenization.  The \photoz\
test also tells us that {\synmag}s have not introduced strong systematic offsets (greater than 0.05 mag) between filters because
these would decrease the \photoz\ precision.  The more general question of whether {\synmag}s or any matching technique introduces
biases in the observed colors requires defining what is meant by ``correct'' colors, against which biases could be measured.  As
we discuss in Section \ref{discussion}, no single definition is appropriate and this argues for adopting different
approaches tied to specific scientific goals.

\begin{figure*}
\plotone{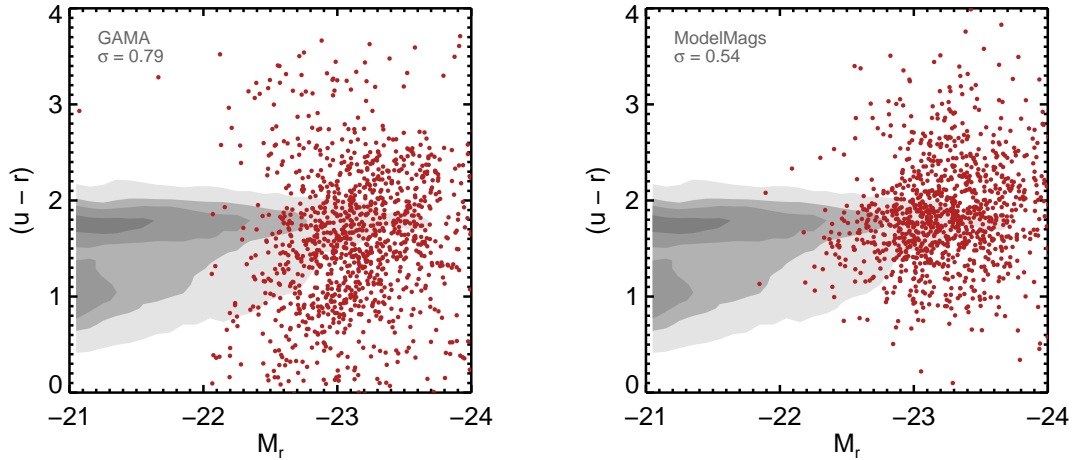}
\caption{Restframe $(u-r)$ color-magnitude diagrams of (mostly) red-sequence CMASS galaxies from the BOSS survey. As with Figure
  \ref{fig:redseq_rj}, the observed
  scatter for the GAMA FWHM-homogenized matched-aperture photometry is larger than for SDSS ModelMags, demonstrating the utility
  of profile-fitting in this context.
  Contours indicate the location of much less luminous COSMOS galaxies in the same redshift range.
  \label{fig:redseq_ur}}
\end{figure*}

\subsection{Why {\synmag}s Perform Well}

A comparison of the red-sequence colors (Figure \ref{fig:redseq_rj}) and photometric redshifts (Figure \ref{fig:photoz_all})
suggests that sets of photometry constructed with {\synmag}s perform better than FWHM-homogenization matched-aperture photometry.
Putting aside the possibility that the GAMA matched-aperture photometry may improve in future releases, it is worth asking why
{\synmag}s may perform better in these tests.  Two explanations are apparent.  To begin with, {\synmag}s do not require a
degradation of the image quality to the lowest-common-denominator PSF.  For many data sets, this is typically FWHM$=$2\arcsec,
which is on the same order as both the size of distant galaxies and the aperture size over which their flux is measured.  The
benefit of not degrading the PSF is demonstrated by the improved scatter of the \synmag\ $(r-J)$ colors.

\begin{figure*}
\plotone{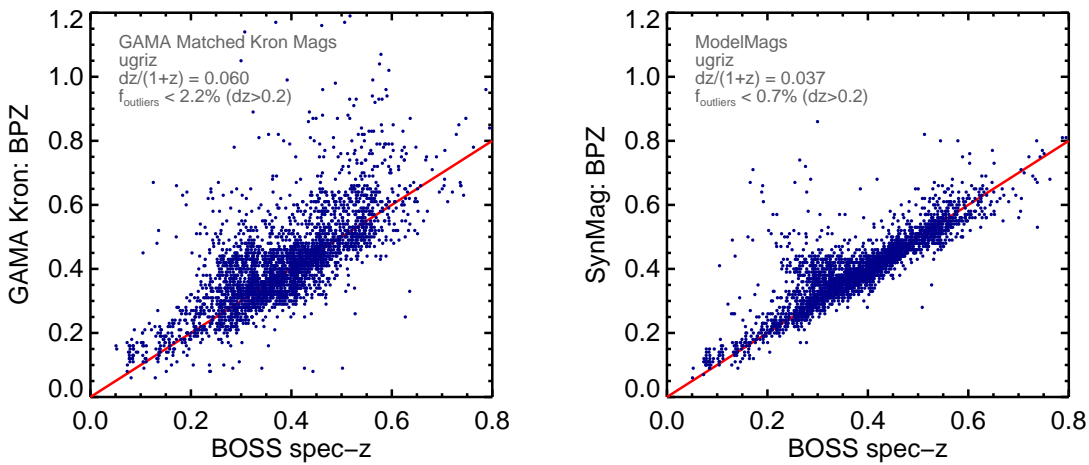}
\caption{The same photometric redshift comparison as in Figure \ref{fig:photoz_all}, but restricting only to SDSS $ugriz$
  photometry.  \synmag\ colors in this case are equivalent to SDSS ModelMags. 
  \label{fig:photoz_ugriz}}
\end{figure*}

The second explanation is that we have constructed a set of photometry that takes advantage of profile-fitting photometry when
available.  The \photozs\ presented above are based on a mix of SDSS ModelMag colors from profile-fitting in the optical and
\synmag\ colors defined by aperture photometry from UKIDSS in the near-IR.  We can examine the value of the profile-fitting colors
alone by ignoring the near-IR {\synmag}s for the moment and comparing colors and \photozs\ between GAMA and SDSS in the optical
only.  Figure \ref{fig:redseq_ur} presents the restframe $(u-r)$ color-magnitude diagrams derived for CMASS galaxies from GAMA
(left panel) compared to the SDSS profile-fitting ModelMags (right panel), and Figure \ref{fig:photoz_ugriz} shows the same test
with $ugriz$ \photozs.  Both show significantly reduced scatter from the use of ModelMags\footnote{It is perhaps surprising that
  the loss of near-IR photometry does not degrade the \photozs.  In fact, for the ModelMags, the \photozs\ significantly {\em
    improve} without the near-IR.  This is in large part a result of uncertainties in restframe near-IR templates, a widespread
  problem that has motivated template error weighting in recent \photoz\ codes \citep[e.g.,][]{brammer08}.}.  The {\synmag}s
themselves cannot claim credit for the reduced scatter in this case, but it is clearly important to make use of profile-fitting photometry if
it is available.  It is likely, for example, that a mix of ModelMags in the optical and FWHM-homogenized colors in the near-IR would
yield tighter colors and \photozs\ than FWHM-homogenized colors in all bands.

\section{Conclusions}\label{discussion}

We have presented a technique for using catalog-level data products to match the photometry of disparate imaging data sets.  Our
{\synmag} approach uses multi-Gaussian expansions to represent both intrinsic profiles and the PSF, allowing for extremely fast
convolutions.  The result is a tool that can be used to quickly obtain matched photometry for large data sets without re-analyzing
pixel data.

We find that {\synmag}s yield colors and derived \photozs\ with scatter that is as good if not better than FWHM-homogenized
aperture photometry.  We have also shown that the resulting scatter is minimized by constructing photometric data sets that employ
profile-fitting colors when available.  These tests, however, {\em assume} that the ``best'' color estimators minimize the
scatter.  This may not be the case, especially given that most galaxies are 2-component bulge$+$disk systems, often with
different colors for each component.  In fact, the choice of aperture size or assumed profile shape should depend on the goal for
which the photometry is being analyzed.  Such ``application-oriented photometry'' would suggest, for example, that for \photozs,
fitting profiles that emphasizes the typically redder light from the bulge would strengthen the 4000\AA\ break in the derived SED
and yield tighter \photoz\ estimates.  The same profile shape would not be optimal for studying the SEDs of galaxy disks, however.

Applying {\synmag}s to future survey data would benefit from having some specific measurements recorded in publicly available
catalogs.  Even if profile-fitting photometry is performed, this basic information is valuable for cross-checks, troubleshooting,
and the application of {\synmag}s to future overlapping data.  First, it would be useful to have aperture photometry performed in a series of
circular apertures that range from roughly one-half of the typical PSF FWHM to $R = 6$--7\arcsec\ (integer values like radii of
2\arcsec, 3\arcsec, 4\arcsec, etc., are common).  The $R = 6$\arcsec\ corresponds to $\sim$11 kpc/$h_{70}$ at $z=0.1$, sufficient
for galaxy photometry at $z > 0.1$.  Several larger apertures up to $R = 20$\arcsec\ would provide adequate sampling for most of
the largest galaxies in SDSS ($z < 0.1$).  No corrections or adjustments (e.g., PSF corrections) should be applied to the aperture
fluxes beyond background subtraction.  Reliable uncertainties on each flux measurement are also valuable.  Special care
is needed for smaller apertures that subtend only a few pixels, otherwise it may be better to remove small aperture measurements
from the catalogs.  It is also useful to consider the definition of null values for measured magnitudes.  The user should be able to
distinguish between the case when no flux was detected (e.g., $m = 99$) and when imaging for a source in a particular band is
missing (e.g., $m = -99$), either because it was not observed or was somehow compromised.  

{\synmag}s also require information about the PSF.  Ideally, each catalog entry would provide a way to determine the PSF shape at
the location of the measured object given the conditions under which it was observed.  Often a 2-component Gaussian description is
sufficient although the modeled shape could be more sophisticated and could be modeled as a function of both detector position and
observing conditions or exposure number.  The position angle of off-axis or non-axisymmetric components also needs to be recorded.

\section{Acknowledgments}
We would like to thank Masahiro Takada, Sogo Mineo, Chiake Higake, and Daniel Thomas for useful discussions that provided new
insight on topics in this paper.  This work was supported by a Kakenhi Grant-in-Aid for Scientific Research 24740119 from the Japan
society for the Promotion of Science.  Further support comes from the World Premier International Research Center Initiative (WPI
Initiative), MEXT, Japan.  Funding for SDSS-III has been provided by the Alfred P. Sloan Foundation, the Participating
Institutions, the National Science Foundation, and the U.S. Department of Energy Office of Science. The SDSS-III web site is {\tt
  http://www.sdss3.org}.

SDSS-III is managed by the Astrophysical Research Consortium for the Participating Institutions of the SDSS-III Collaboration
including the University of Arizona, the Brazilian Participation Group, Brookhaven National Laboratory, University of Cambridge,
Carnegie Mellon University, University of Florida, the French Participation Group, the German Participation Group, Harvard
University, the Instituto de Astrofisica de Canarias, the Michigan State/Notre Dame/JINA Participation Group, Johns Hopkins
University, Lawrence Berkeley National Laboratory, Max Planck Institute for Astrophysics, Max Planck Institute for
Extraterrestrial Physics, New Mexico State University, New York University, Ohio State University, Pennsylvania State University,
University of Portsmouth, Princeton University, the Spanish Participation Group, University of Tokyo, University of Utah,
Vanderbilt University, University of Virginia, University of Washington, and Yale University.

\begin{appendix}
\section{Cookbook: Using the {\sc synmag} software}\label{cookbook}

This appendix presents instructions for using the {\synmag} software\footnote{{\tt
    http://member.ipmu.jp/kevin.bundy/synmag}}.  We describe the key steps in the process and point to the routine {\tt synmag\_sdss} for the
  specific case of matching to SDSS data.

Before using the \synmag\ routines, some photometric information from the data sets involved is required.  This should include parameters from
profile fitting in at least one band, aperture photometry results for the ``target'' data set where profile fits are not available, and
seeing estimates for this data set.  With these information in hand, the procedure is as follows:

\begin{enumerate}

\item {\it Decompose the intrinsic profile shape into an MGE.}  This only needs to be done once for each profile shape that is
  used.  For standard de Vaucouleurs and exponential profiles, or their SDSS variants, the Gaussian coefficients can be read off
  of Table \ref{table:mge}.  Assuming more generally that the profile is a \sersic, the {\tt mgsersic} routine can be used to
  determine the MGE.  It utilizes a 1D fitting routine in the {\sc mge\_fit\_sectors} IDL package presented in
  \citet{cappellari02}.  As discussed in Section \ref{decomposition}, however, a least-squares MGE fit to the 2D profile provides a better
  approximation \citep[see][]{hogg12}.  The PSF of the target catalog can also be decomposed in this way if desired.

\item {\it Compute synthetic photometry.} Use {\tt synmag} to compute the synthetic aperture photometry in the profile-defining
  band for the PSF and aperture size measured in the target catalog.  Multiple apertures can be requested and the PSF parameters
  can be specified separately for each source to account for variations across the image or from differing observing epochs.

\item {\it Assemble a matched set of photometry.} Repeat the {\tt synmag} procedure for each filter band in the target catalog to account for PSF variations and
  missing data from one band to the next.  Then subtract the resulting colors from the total flux measured in the profile-defining
  band in order to construct a set of matched flux total flux measurements (i.e., $m_{i,\rm tot} = m_{p,\rm tot} - C_{pi}$, see
  Section \ref{method:overview}).  

\end{enumerate}

The PSF-convolved MGEs can also be returned by {\tt synmag}.  The routine {\tt mgestr} computes the effective radius and axis
ratio of any MGE (see Section \ref{sec:mgestr}).  

\end{appendix}

\bibliographystyle{apj}
\bibliography{references}

\end{document}